# Model-Free Market Risk Hedging Using Crowding Networks

Vadim Zlotnikov[1], Jiayu Liu, Igor Halperin, Fei He, Lisa Huang[2]

Fidelity Investments, Boston MA 02210[3]

{Vadim.Zlotnikov, Jiayu.Liu, Igor.Halperin, Fei.He, Lisa.Huang}@fmr.com

June 12, 2023

**Abstract**

Crowding is widely regarded as one of the most important risk factors in designing portfolio strategies. In this paper, we analyze stock crowding using network analysis of fund holdings, which is used to compute crowding scores for stocks. These scores are used to construct costless long-short portfolios, computed in a distribution-free (model-free) way and without using any numerical optimization, with desirable properties of hedge portfolios. More specifically, these long-short portfolios provide protection for both small and large market price fluctuations, due to their negative correlation with the market and positive convexity as a function of market returns. By adding our long-short portfolio to a baseline portfolio such as a traditional 60/40 portfolio, our method provides an alternative way to hedge portfolio risk including tail risk, which does not require costly option-based strategies or complex numerical optimization. The total cost of such hedging amounts to the total cost of rebalancing the hedge portfolio.

## 1   Introduction

Market dynamics are driven by collective actions of market participants, who respond to new information by buying and selling assets and reallocating capital across assets and asset classes. Allocation activity by market participants is greater than new investment inflows, and largely determines pricing of capital assets over short-to-medium term horizons. Since 2003 we have seen a significant increase in correlation of different assets (e.g. equity and bonds across different countries), as well emergence of "thematic" clusters with much higher pair-wise correlations. There are many possible reasons for these observations, some of which are external (e.g. uniformity of central banks' policies, globalization of business and pools of capital), or common of investor behavior (e.g. ubiquitous information access, reliance on similar heuristics and risk/portfolio construction models, identical data sources for back testing etc). We expect many of these factors to persist, which would complicate the task of diversification, as well as result in occasional, sharp market drawdowns as commonly held beliefs are reverse. Viewed through another lens, we are seeing a significant rise in the degree of "herding or crowding" among investors with similar time horizons, which dramatically increases endogenous risk associated with investors' positioning. Therefore, understanding and defining 'crowding' is vital for investment management.

Crowding is a term widely used in industry, albeit not always in the same way or using the same methodology. Crowding is often mentioned as a single most important risk factor beyond factors captured by classical equity cross-sectional models such as the Fama-French or Carhart models. Various approaches to crowding vary from purely prescriptive ex-post methods to predictive (ex-ante) methods that try to incorporate crowding within a forward-looking perspective, with an objective to use it for asset allocation decisions. One approach to crowding is based on analysis of crowding at the factor level rather than on a single stock level, and correlation patterns of stock returns, see e.g. [1] and references therein. Alternatively, there are holding-based approaches to crowding, see e.g. [2] for a graph-based method. Our approach is similarly based on a network analysis of fund holdings, though it is different in methodology. We note that holding-based graph analysis of mutual funds was also considered in [3], though not in the context of stock crowding which is the focus of this research.

---

[1] Corresponding authors
[2] Corresponding authors
[3] Opinions presented in this paper are authors' only, and not necessarily of their employer. The standard disclaimer applies.

The AB research from 2013 [4,5] considers crowding for international markets. The analysis targeted the design of an overall crowding metric that incorporates 5 individual crowding indicators: percentage buy ratings, adjusted over-weights by active managers, institutional trade persistence, long-term price momentum, and aggressive expectations with implied achievability. The analysis in [4,5] suggested that crowded trades are characterized by the following traits:

- enhanced forward-looking volatility relatively to non-crowded stocks

- crowded stocks tend to have negative quadratic beta [4][4], which suggests such stocks may underperform during periods of high market volatility

- strong negative skew in reaction to earning revisions

- crowding may not predict underperformance at the stock level, but rather suggests underperformance at the sector and region level

- Crowded stocks tend to have a low correlation with the market, but enhanced correlations between themselves. This may translate into increased risk at a portfolio level, if crowding is not included in a portfolio risk model.

In Zlotnikov *et. al.* 2022 [6] a more refined holdings-based crowding signal was computed for all stocks in Russell 1000 using Thompson-Reuters data from 2000 to March 2021 for holdings of all actively managed mutual funds and hedge funds in 13F reports. The crowding signals were further used to construct a diversified portfolio that has desired characteristics during periods of market downturns.

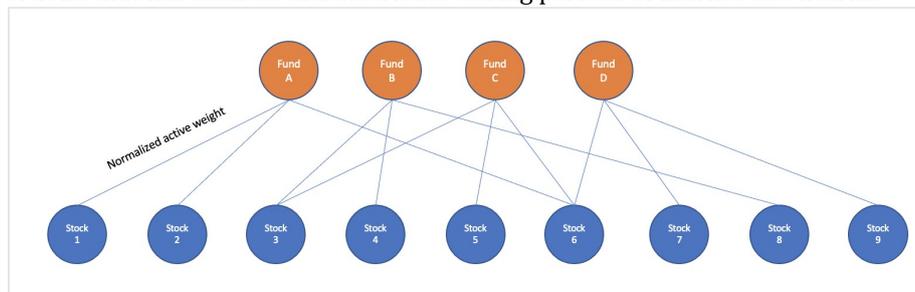

Figure 1: A bipartite graph is constructed using the holdings data consisting of fund nodes and stock nodes. The links between them are the normalized active weight of the stock in the fund. A bipartite graph is a type of graph in which the nodes can be divided into two disjoint sets. There are no links between nodes of the same type.

In this paper, we take a different approach and construct a different crowding signal that uses one of the five crowding indicators mentioned above, namely adjusted over-weights by active managers are used. This can be viewed as a simplification of the original approach of [4]. In addition to this simplification, there is also a technical innovation in our new approach. More specifically, we build a crowding signal using graph features. The intuition behind why we think graphs are appropriate for the construction of crowding signal is because 'crowding' is conceptually a network effect. Previous work using holding base approaches fails to capture this given that it effectively operated at the '1-hop' level.

---

[4] Quadratic beta is a regression coefficient for the square of market returns in a regression of stock returns on the market return and its square.

We construct a bipartite graph representing all stocks and all funds, where capitalization-adjusted active weights are used as edges between funds and stocks. We then use the centrality measures calculated from the holdings graph to represent the crowdedness of stocks. We then demonstrate how our proposed graph-based crowding signal computed using eigenvector centrality can be used to produce a crowding-based long-short hedge portfolio which is negatively correlated with the market, and provides a downside protection during periods of market downturns.

## 2 Methodology

### 2.1 A graph based crowding signal

We construct a holdings-based crowding signal using holdings data for actively managed US funds covering the period from 2014 to 2022 from Morningstar. This data is first used to calculate the active weight for each stock. To remove a potential bias introduced by market capitalizations, we calculate normalized active weights by dividing active weights by the logs of market capitalization.

For each quarter, we construct a graph based on the holdings data. Figure 1 shows the bipartite graph we constructed. It consists of the fund nodes and stock nodes. The links between funds and stocks are the normalized active weight of the stock in the fund.

To define appropriate centrality measures, we split the graph into an overweight graph and an underweight graph. That is, for the overweight graph we only keep edges that correspond to overweights, and do symmetrically for the underweight graph. To reduce potential numerical noise in our procedure, we only keep those links whose weight is above a median threshold. We then calculate different centrality measures (degree centrality, weighted degree centrality and eigenvector centrality) of each stock in both the overweight and underweight graphs. The crowding signal is then defined as the difference between the centrality in the overweight graph and that in the underweight graph. The intuition is that the more crowded a stock is, we expect it to be more important in the overweight graph and less important in the underweight graph, because a lot of funds will be over-weighing it instead of under-weighing it.

### 2.2 Back testing the graph base crowding signal

At each quarter, we create a long/short portfolio that uses the crowding signal to go long for uncrowded stocks, and short crowded stocks. To control potential factor biases in our signal, we use a proprietary optimizer to limit factors exposure between -0.02 and 0.02 for the following five Barra factors: beta, growth, momentum, volatility and size. We also limit the number of stocks held on each side of our long/short portfolio by 100.

Considering that fund holdings data is only available about 45 days after the quarter end, we construct the portfolios 2 months after the quarter end. We observe the returns for the next 1, 3, 6, and 12 months. We then calculate the following metrics for the returns series.

- mean of the returns

- skewness of the returns

- correlation with the market returns (Russell 1000)

- quadratic beta, which is the loading on market return squared [4]

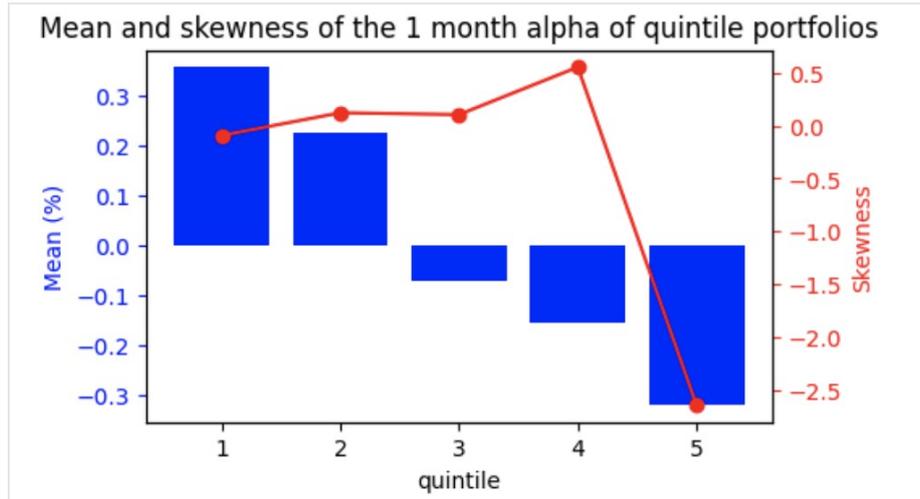

Figure 2: Mean and skewness of the 1 month alpha of quintile portfolios based on eigenvector centrality (2014-2022). We divide the stocks in our universe into quintiles based on eigenvector centrality, with quintile 1 having the least crowded stocks, and quintile 5 having the most crowded stocks. We equal weight the stocks in each quintile, and then observe the return of each quintile over the next month relative to benchmark, which consists of all the stocks equal-weighted. On average the most crowded quintile underperforms the less crowded quintiles with a strong negative skewness. We build a 2 month lag between the actual holdings date and the portfolio construction date. The 1 month return is observed after the portfolio construction date.

## 3 Results

### 3.1 Crowding signal from eigenvector centrality

We divide the stocks in our universe into quintiles based on the difference of eigenvector centrality in the overweight graph and underweight graph, with quintile 1 having the least crowded stocks, and quintile 5 having the most crowded stocks. We use equal weights for all stocks in each quintile, and then observe the returns of these quintile portfolios relative to benchmark, which consists of all the stocks equal-weighted. Figure 2 shows the mean and skewness of the 1-month alpha computed in this way for all quintile portfolios. We can see that on average the most crowded quintile clearly underperforms the less crowded quintiles. It also has a strong negative skewness, unlike the other quintiles which have close to zero or slightly positive skewness. This behavior is consistent with the conclusion of [4].

Figure 3 shows the factor tilts of the quintile portfolios relative to benchmark. The five factors shown here are beta, growth, momentum, volatility and size from Barra. We can see that the most crowded and least crowded quintiles have the smallest sized stocks, and therefore, higher volatility. We also observe that quintile 1 (least crowded quintile) has constant negative exposure to momentum, which is expected.

We can see from the above simple quintile analysis that crowded stocks have certain undesirable features compared to uncrowded stocks, like negative mean return over 1 month after the portfolio construction date and negative skewness. To remove the impact from factor tilts, we use an optimizer to create a long/short portfolio that is factor neutral while minimizing the crowding signal. We also limit the number of stocks on each side to 100.

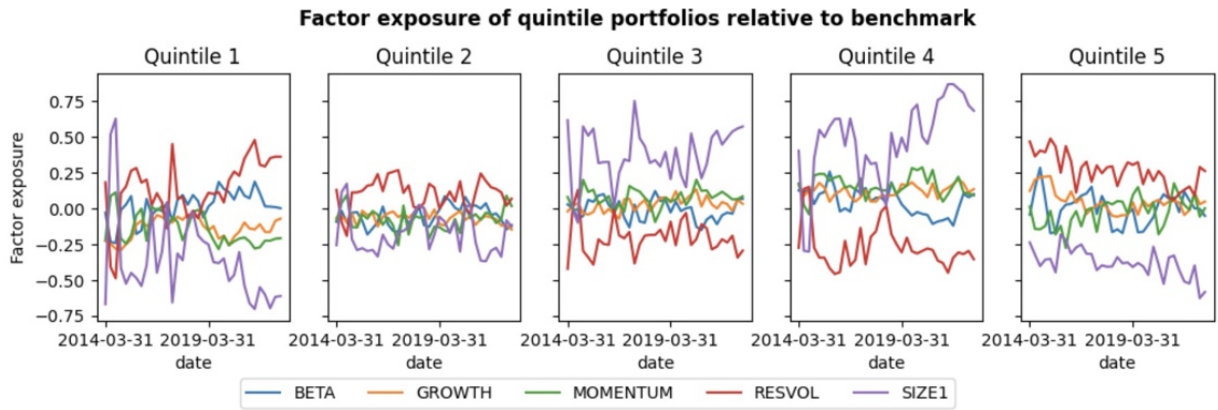

Figure 3: Factor exposure of quintile portfolios relative to benchmark (2014-2022). The five factors shown here are: beta, growth, momentum, volatility, and size as defined by Barra. The most crowded and least crowded quintiles have the smallest size stocks, and also higher volatility. The least crowded quintile also has negative exposure to momentum.

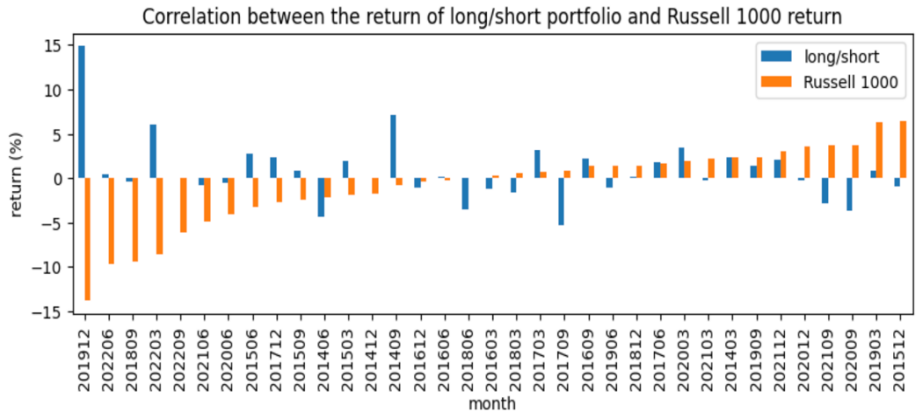

Figure 4: Correlation between the 1 month return of long/short portfolio and Russell 1000 return, sorted by Russell 1000 monthly return (2014-2022). The long/short portfolio is constructed from an optimizer that minimizes the crowding signal while keeping the factor tilts close to neutral. The returns of the long/short portfolio and Russell 1000 appear to be negatively correlated.

Figure 4 shows the 1 month returns of long/short portfolio and Russell 1000, sorted by Russell 1000 monthly return. We can see that their returns seem negatively correlated. The correlation is -0.47. We note that when Russell 1000 suffered negative returns, the long/short portfolio fared much better. We also note that when Russell 1000 had positive returns, the long/short portfolio either had positive return, or had slightly negative return. This is a desirable property to have for a hedge portfolio that provides down-side protection.

Figure 5 shows the scatter plot between long/short portfolio returns and Russell 1000 returns. The red line is the quadratic regression line. We can see that the curve has a convex shape, corresponding to a positive coefficient in front of the second order term. This means that when there is out-sized movement in the market return, the long/short portfolio cushions the return by contributing positively.

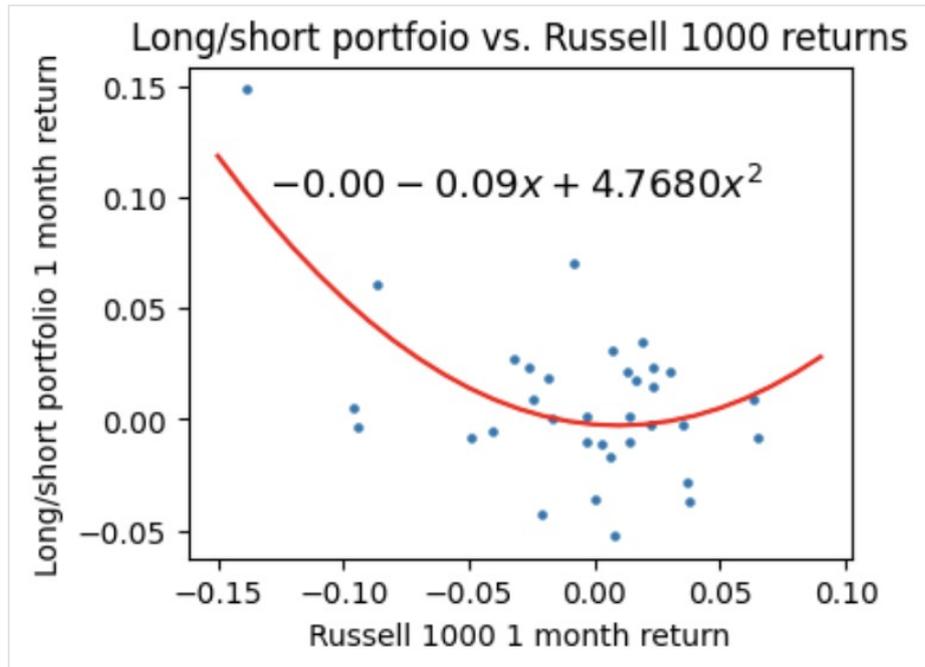

Figure 5: Scatter plot between the 1 month long/short portfolio returns and Russell 1000 returns (2014-2022). The long/short portfolio is constructed from an optimizer that minimizes the crowding signal while keeping the factor tilts close to neutral. The red line shows the quadratic regression line, which has a convex shape, indicating that the long/short portfolio contributes positively when there is out-sized movement in the market.

In addition, the long/short portfolio has a positive average 1 month return of 0.78% and a positive skewness of 1.76 over the time period in our research. So overall, this portfolio has negative correlation with the market, positive skewness, and positive quadratic beta, which are all desirable properties of a hedge portfolio. This insurance does not cost anything, or may even provide some positive alpha.

## 3.2 Comparison with other crowding signals

How does this graph-based eigenvector centrality signal compare with other simple signals based on counting-based methods? Here we compare with the following signals:

- Degree centrality: This is also graph based. However, it simply counts how many immediate neighbors a node has. Unlike eigenvector centrality, neighbors more than 1-hop away do not impact this metric.

- Weighted degree centrality: It also counts how many immediate neighbors a node has, but each neighbor contributes unequally. It is weighted by the link value. Again, neighbors more than 1-hop away do not impact it.

- The crowding signal proposed in [6], which consists of aggressive ownership, ownership concentration, high-risk ownership, active tilt, aggressive accumulation, and consistent accumulation.

Figure 6 compares these different signals on mean return, skewness, correlation with the market and quadratic beta. We can see that eigenvector centrality performs the best in all these metrics.

|                            | mean   | skewness | correlation with market | quadratic beta |
|----------------------------|--------|----------|-------------------------|----------------|
| Degree centrality          | 0.47%  | 0.85     | -0.39                   | 0.76           |
| Weighted degree centrality | 0.45%  | 0.65     | -0.35                   | 2.71           |
| Eigenvector centrality     | 0.77%  | 1.84     | -0.49                   | 5.01           |
| Signal from [6]            | 0.15%  | -0.24    | 0.09                    | 0.89           |

Figure 6: Comparison of crowding signals constructed from different methods on mean return, skewness, correlation with the market and quadratic beta (2014-2022). Degree centrality is a measure that simply counts how many immediate neighbors a node has. Weighted degree centrality also counts the number of immediate neighbors, but each neighbor contributes unequally according to the link value. The signal from 6 measures is defined in [6], which consists of aggressive ownership, ownership concentration, high-risk ownership, active tilt, aggressive accumulation, and consistent accumulation. The eigenvector centrality measure, which considers ownership from multiple hops, performs the best in all these metrics.

# 4    Conclusion

While crowding is widely recognized as an important risk driver alongside classical factors such as Fama-French or Carhart, a technical implementation of this concept had remained somewhat elusive in the literature, without producing any commonly accepted framework. In this paper, we contribute to the literature by addressing the crowding problem using graph analysis. Graph analysis is currently not widely used in investment management even though many problems lend themselves to a representation as a graph. Using graphs to understand crowding is visually intuitive because crowded nodes are the ones that have many links pointing to them. Using graph eigenvector centrality metrics, we compute stocks' crowding scores, which are then used to form a long-short portfolio that possess some attractive characteristics for our objectives, such as positive convexity and negative correlation with the market.

Typically, to get convexity in a portfolio, we would need to use derivative instruments. Though adding derivatives to stock portfolios to produce convex payoff profile and hence provide a downside protection is a straightforward and commonly used practice, it has upfront costs that may be steep at times. In contrast, here we can use the intuition of crowding to create a costless long-short portfolio that is negatively correlated with the market, has positive skewness in return and has positive convexity. It can serve as a hedge, especially during volatile periods. In addition, it has positive overall alpha in the back tests during the periods that were tested. As the upfront cost of our hedge is zero, the total hedging cost amounts to total rebalancing costs of the hedge portfolio. Using our proposed long-short portfolio as a hedge of market risk including both small and large market fluctuations can therefore be viewed as an attractive alternative to costly option-based strategies that are often employed to mitigate portfolio tail risk.

Our construction of the long-short hedging portfolio is ``model-free'', or more accurately distribution-free, as it relies only on the holding information for funds, and not on any probabilistic assumptions about the future market performance. (We do rely though on standard portfolio risk models when we decide on an optimal mixing of our long-short portfolio with the baseline portfolio.). Moreover, construction of our long-short portfolio does not rely on any optimization and uses only linear algebra.

While in this paper we simply use classical graph features like eigenvector centrality, more advanced graph machine learning methodology can also be used to study crowding. In particular, machine learning can be used to produce nowcasting for fund holdings, which could be valuable given that fund holding information is available only with a lag. Furthermore, we could use dynamic graph models to model the time evolution of funds' holdings. With such models, forecasting (or nowcasting) of funds' holdings amounts to the problem of link prediction. Modern tools such as graph neural networks (GNN) for dynamic graphs (see e.g. [7]) thus appear a promising direction for future research in these directions.

# References

[1] Nick Baltas, The Impact of Crowding in Alternative Risk Premia Investing. *Financial Analysts Journal*, 75(3), pp.89-104, 2019.

[2] van Kralingen, M., Garlaschelli, D., Scholtus, K. and van Lelyveld, I., Crowded Trades, Market Clustering, and Price Instability. *Entropy*, *23*(3), p.336, 2021.

[3] Vipul Satone, Dhruv Desai and Dhagash Mehta, Fund2Vec: Mutual Funds Similarity using Graph Learning, https://arxiv.org/abs/2106.129872021, 2021.

[4] Vadim Zlotnikov, Quantitative Research: Expanding Analysis of Crowding to Global Markets - Part 1, 2013.

[5] Vadim Zlotnikov, Quantitative Research: Comprehensive Framework for Crowding in Global Markets - Part 2, 2013.

[6] Vadim Zlotnikov, Mikhail Stukalo, Igor Halperin, Lisa Huang, Cathy Pena, Questioning the Wisdom of the Crowds to Design the Portfolio Diversification Strategies, *The Journal of Alternative Investments,* DOI 10.3905/jai.2022.1.178, 26 November 2022.

[7] Joakim Skarding, Bogdan Gabrys, Katarzyna Musial, Foundations and Modeling of Dynamic Networks Using Dynamic Graph Neural Networks: A Survey, *IEEE Access*, 2021 (https://ieeexplore.ieee.org/stamp/stamp.jsp?arnumber=9439502).